\documentclass [preprint]{aastex}

\shorttitle{Classic FR-II BAL QSO J1016+5209}
\slugcomment{Submitted to the {\em Astrophysical Journal}}
\begin{document}

\def\roma#1{\ifmmode{#1}\else{$#1$}\fi} 
\def\extra#1{\roma{\phantom{\rm#1}}}
\def\kms{\roma{\,\rm km\,s^{-1}\,}}                    
\def\kmsmpc{\roma{\rm\,km\,s^{-1}\,Mpc^{-1}}}                       
\newcommand{\Lya}{Ly$\alpha$}
\newcommand{\MgII}{\ion{Mg}{2}}
\newcommand{\MgI}{\ion{Mg}{1}}
\newcommand{\CaI}{\ion{Ca}{1}}
\newcommand{\CaII}{\ion{Ca}{2}}
\newcommand{\CIII}{\ion{C}{3}]}
\newcommand{\CIV}{\ion{C}{4}}
\newcommand{\FeI}{\ion{Fe}{1}}
\newcommand{\FeII}{\ion{Fe}{2}}
\newcommand{\FeIII}{\ion{Fe}{3}}
\newcommand{\SiIII}{\ion{Si}{3}]}
\newcommand{\SiIV}{\ion{Si}{4}}
\newcommand{\AlIII}{\ion{Al}{3}}

\title {\bf Discovery of a Classic FR-II Broad Absorption Line Quasar from the
FIRST Survey}

\author{
Michael~D.~Gregg\altaffilmark{1,2},
Robert~H.~Becker\altaffilmark{1,2},
Michael~S.~Brotherton\altaffilmark{2,3},
Sally~A.~Laurent-Muehleisen\altaffilmark{1,2},
Mark~Lacy\altaffilmark{2,1},
Richard~L.~White\altaffilmark{4}
}

\altaffiltext{1}{Physics Dept., University of California, Davis, CA
95616, gregg,bob,slauren,mlacy@igpp.ucllnl.org}
\altaffiltext{2}{Institute for Geophysics and Planetary Physics, L-413
Lawrence Livermore National Laboratory, 7000 East Avenue, Livermore, CA 94550}
\altaffiltext{3}{National Optical
Astronomy Observatory, 950 North Cherry Avenue, Tucson, AZ 85726 mbrother@noao.edu}
\altaffiltext{4}{Space Telescope Science Institute, 3700 San Martin
Drive, Baltimore, MD 21218, rlw@stsci.edu}

\begin{abstract}

We have discovered a remarkable quasar, FIRST J101614.3+520916, whose
optical spectrum shows unambiguous broad absorption features while its
double-lobed radio morphology and luminosity clearly indicate a
classic Fanaroff-Riley Type II radio source.
Its radio luminosity places it at the extreme of the recently
established class of radio-loud broad absorption line quasars (Becker
et al.\ 1997, 2000; Brotherton et al.\ 1998).  Because of its hybrid
nature, we speculate that FIRST J101614.3+520916 is a typical FR-II
quasar which has been rejuvenated as a broad absorption line (BAL) quasar
with a Compact Steep Spectrum core.
The direction of the jet axis of FIRST J101614.3+520916 can be
estimated from its radio structure and optical brightness, indicating
that we are viewing the system at a viewing angle of $\gtrsim
40\arcdeg$.  The position angles of the radio jet and optical
polarization are not well-aligned, differing by $\sim 20\arcdeg -
30\arcdeg$.  When combined with the evidence presented by Becker et
al.\ (2000) for a sample of 29 BAL quasars showing that at least some
BAL quasars are viewed along the jet axis, the implication is that no
preferred viewing orientation is necessary to observe BAL systems in a
quasar's spectrum.  This, and the probable young nature of compact
steep spectrum sources, leads naturally to the alternate hypothesis
that BALs are an early stage in the lives of quasars.

\end{abstract}

\section {Introduction}

It was long believed that a quasar could not be radio-loud and
simultaneously have broad absorption lines in its optical spectrum
(Stocke et al.\ 1992; Hamann, Korista, \& Morris 1993; others).  Over
the past several years, the existence of quasars exhibiting both
properties has been firmly established (Becker et al.\ 1997;
Brotherton et al.\ 1998; White et al.\ 2000; Becker et al.\ 2000).  We
report here the most striking example to date of a radio-loud BAL
quasar at z = 2.455, FIRST~J101614.3+520916, which is not only
extremely radio-loud but has a classic Fanaroff-Riley Type II (FR-II;
Fanaroff \& Riley, 1974) morphology with very bright radio lobes.

About 10\% of the known quasars exhibit BAL outflows in their spectra
(though the actual fraction depends heavily on how samples are
selected, see Becker et al.\ 2000).
This has been generally interpreted as BAL quasars being ordinary
quasars viewed along a line of sight grazing the optically thick torus
which surrounds the massive black hole engine (Weymann et al.\ 1991).
The broad, highly blueshifted absorption is postulated to arise from
clouds which are evaporated from the torus and accelerated outward by
radiation pressure.  In this picture, however, the apparent dichotomy
of strong radio emission and the BAL phenomenon is extremely puzzling.
There is no obvious reason in the orientation picture for radio
emission to be suppressed, especially that from extended radio lobes.
Additionally, typical BAL quasars have emission line and continuum
spectral properties indistinguishable from ordinary quasars, in both
the optical and radio (Stocke et al.\ 1992; Weymann et al.\ 1991;
Barvainis \& Lonsdale 1997).  A number of possible explanations for
the lack of radio-loud BALs -- free-free absorption, frustrated jets,
small scale structure in the accretion region -- have been suggested
(Stocke et al.\ 1992; Begelman, de~Kool, \& Sikora 1991; Boroson,
Persson, \& Oke 1985), but all are problematic.  So perhaps it is not
surprising that radio-loud BALs have finally been found.

This shift in understanding is being driven by studies which are
probing new parts of parameter space: the FIRST Bright Quasar survey
(FBQS; Gregg et al.\ 1996; White et al.\ 2000) relies on
unprecedentedly faint radio fluxes of 1~mJy for candidate selection.
The only other significant sample of radio-loud BALs, that of
Brotherton et al.\ (1998) which found 5, also used a faint radio
flux-limited sample (down to 2.5~mJy) selected from the NRAO VLA Sky
Survey (NVSS, Condon et al.\ 1998) to select quasar candidates.  These
studies are the first which are radio-sensitive enough to probe the
transition region between radio-quiet and radio-loud quasars with good
statistics.  The radio properties of the BAL quasars, both radio-quiet
and loud, present serious challenges to the current understanding of
the quasar phenomenon.

FIRST~J101614.3+520916 (hereafter J1016+5209) is a BAL quasar which we
have found in an ongoing project to extend the FBQS from a limiting
magnitude of E=17.8 to E=19.  With the discovery of this object, any
remaining uncertainty over whether a BAL quasar can be radio-loud is
completely removed.  Further, because of its FR-II morphology, some
constraints can be placed on the viewing angle and orientation of the
system.  The properties of J1016+5209, taken together with the general
results of the BAL quasar study of Becker et al.\ (2000), may call for
a fundamental alteration of the commonly accepted
unification-by-orientation scheme, or at least how the BAL phenomenon
fits into the general quasar model.

The only other object known which may be similar to J1016+5209 in
combining the characteristics of BALs with FR-II radio morphology is
the z=0.240 quasar PKS~1004+13 (Wills, Brandt, \& Laor 1999).
PKS~1004+13 is not as extreme in its radio-loudness nor in the
strength of its broad absorption (about which there is some remaining
doubt due to the low S/N of the defining IUE spectrum), but if real,
there are now two members of this hybrid class.

\section {Observations}

A POSS-I stellar counterpart with E and O magnitudes of 18.6 and 20.2,
respectively, lies within 0\farcs25 of the radio source
J1016+5209.\footnote{These magnitudes are on the ``corrected'' APM
system of White et al.\ (2000); photographic O and E are comparable to
the more familiar B and R bands.  Galactic reddening in this direction
is insignificant, A$_V = 0.005$.}  A 9\AA\ resolution optical spectrum
was obtained using the Low Resolution Imaging Spectrograph (LRIS, Oke
et al.\ 1995) at the Keck Observatory in December, 1998.  A longer
exposure but lower resolution ($\sim$ 20\AA) LRIS spectrum was taken
at Keck in June, 1999 (Figure~1).  Both spectra clearly show prominent
broad absorption features.  
The emission line redshift is 2.454 based on the
fitted peak of \CIII~1909 which is unaffected by BAL features,
though it may be contaminated by \FeIII, \SiIII, and \AlIII\ emission.  

A second, higher S/N, 9\AA\ resolution spectrum was obtained with LRIS
in November, 1999.  A close look at these data plotted in velocity
space (Figure~2), shows an overall similarity between the \CIV\ and
\SiIV\ BAL systems, though they differ in some details.  Both \CIV\
and \SiIV\ exhibit a very broad system at $\sim -15,000$ \kms and
deeper but more complex absorption system at velocities from 0 to
-8000 \kms.  The broad \CIV\ feature extends from $\sim -8500$ \kms to
velocities of at least -17,200 \kms from the line center, possibly to
-20,000 \kms, at a depth $\gtrsim 10\%$ of the continuum.  Both \SiIV\
and \CIV\ also have significant absorption to the red of the line
centers, by $\sim$ 1000~\kms.  The \CIV\ ($\lambda\lambda$ 1548.2,
1550.8) doublet is separated by only $\sim 500$ \kms while the \SiIV\
($\lambda\lambda$ 1393.8, 1402.8) doublet is separated by a much
greater 1920 \kms; three obvious velocity systems seen in both species
are marked in Figure~2.  The LRIS spectral resolution is comparable to
the redshifted separation of the \CIV\ doublet (8.5\AA), but even the
relatively narrow absorption trough at -6500 \kms\ has a FWHM of
23\AA, about twice as broad as the instrumental resolution convolved
with a single \CIV\ doublet, indicative of broad or multiple
components.  The \SiIV\ doublet is easily split and the widths of the
narrow individual lines are consistent with the instrumental
resolution.  In addition to the narrow features, the general continuum
level in the velocity interval 0 to -10,000 \kms is depressed by very
broad absorption comparable in depth to the higher velocity BAL.

Using the \CIV\ region, we have calculated the ``BALnicity'' index of
J1016+5209 to be 2401 \kms\ (see Appendix A of Weymann et al.\ 1991).
The BALnicity index (BI), though a quantitative measurement, is
sensitive to subjective considerations, particularly continuum
placement.  For J1016+5209, the regions around -9000 and -5000 \kms
(Figure~2) are particularly important; if counted as continuum, then
the BI drops to $\sim 1600$ \kms, based solely on the very broad
absorption extending from -17,200 to -8500 \kms.  Even this lower
value is well within the ranks of other BAL quasars (Weymann et al.\
1991).

As an ultimate test of its BAL nature, we obtained a high resolution
spectrum of J1016+5209 in 2000 April, using the Echelle Spectrograph
and Imager (ESI; Epps \& Miller 1998) on the Keck~II telescope.  With
a 1\arcsec\ slit, the instrument delivers a dispersion of 0.15\AA\ to
0.3\AA\ per pixel over a wavelength range of 3900 to 10900\AA, highly
oversampling the $\sim1.5$\AA\ resolution spectrum.  In Figure~3, we
show the 4800 to 5500\AA\ (restframe 1390 to 1590\AA) region from the
1800s integration ESI spectrum, smoothed with a 9-pixel box,
appropriate for the high oversampling, to improve the S/N.
Overplotted is the LRIS spectrum from Figure~2.  The S/N of the ESI
spectrum is somewhat lower, and there is some additional structure in
the depths of the BAL features, as might be expected from higher
resolution data.  But it is apparent that over the velocity range
-17200 to -1500 \kms\ ($\sim$1460-1540\AA\ rest wavelengths), none of
the broad absorption breaks up into discrete narrow-line systems,
confirming the BAL nature of J1016+5209.  An example of what might be
expected if the BAL regions in J1016+5209 did break up into discrete
narrow-line clouds can be seen in the right panel inset of Figure~3
which shows the striking increase in resolution provided by the ESI
spectrum for the intervening \MgII\ system, a truly narrow-line
absorber.  The only place where the BAL troughs do resolve into
discrete components is at 5330\AA\ (observed), but this is within
$\sim 1400$ \kms\ of the \CIV\ peak and hence does not contribute to
the ``BALnicity'' index.

We have also obtained spectropolarimetry with Keck and LRIS in 2000
January, as part of a broader program to obtain spectropolarimetry for
all of the FBQS BAL quasars.  J1016+5209 is polarized at the 2.5\%
level, rising gradually from less than 2\% at 8000\AA\ (2350\AA\ rest)
to about 3\% at 4200\AA\ (1250\AA\ rest).  The polarization position
angle varies from 85\arcdeg\ to 75\arcdeg\ over the same wavelength
interval.  These polarization characteristics are typical of BAL
quasars (Hines \& Wills 1995; Goodrich \& Miller 1995; Cohen et al.\
1995).  We will discuss the polarization properties of J1016+5209 in
more detail in a future paper.

A contour plot of the FIRST survey 1400~MHz radio image of J1016+5209
is displayed in Figure~3a.  This field has the typical FIRST
\footnote{The FIRST Survey World Wide Web homepage is
http://sundog.stsci.edu} survey image characteristics: 5\arcsec\
resolution and 0.15 mJy RMS noise.  The core radio component
associated with the quasar is marginally resolved, with a deconvolved
size of 4\arcsec, and has a flux density of 6.5 mJy.  It is bracketed
by two bright radio sources of 131 and 39 mJy, both slightly resolved
in the FIRST data.  We interpret these as edge-brightened radio lobes,
making J1016+5209 a classic triple radio source with FR-II morphology
and a total flux density of 177 mJy.  The contours in Figure~3a
suggest that there is a physical connection between the core and the
brighter lobe.  The total angular distance between lobe centers is
45\arcsec\ at a position angle of 146\arcdeg.
The radio source is bright enough to appear in several other radio
surveys.  The NVSS (Condon et al.\ 1998) lists a 20cm flux of 174 mJy
for this object, indicating that FIRST adequately detects all of the
flux and also that the source is probably not highly variable on
timescales of a few years.  The 92~cm WENSS survey (Rengelink et al.\ 1997)
measured a total flux density of 850 mJy.  The source is also
detected in the 6~cm Greenbank (Becker, White, \& Edwards 1991) survey
with a total flux density of 44 mJy.  The WENSS and Greenbank data
yield a global spectral index of $\alpha = -1.1$ (where $S_\nu \propto
\nu^\alpha$), dominated by the bright lobes.

In 1999 July, a 0\farcs25 resolution Very Large Array\footnote{The
Very Large Array is a facility of the National Radio Astronomy
Observatory, operated by Associated Universities, Inc., under
cooperative agreement with the National Science Foundation.} (VLA)
image of J1016+5209 was obtained in the A-configuration at 3.6 cm
wavelength.  The core is just marginally resolved with a fitted flux
density of 1.84~mJy in this new image (RMS of 0.078~mJy); however, the
northwest lobe shows an extended hotspot (Figure~3b) with a flux of
13.8~mJy and deconvolved size of $0\farcs59 \times 0\farcs15$.  These
are lower limits as flux on scales greater than a few arcseconds will
be resolved out.  The position angle of the major axis of the resolved
hotspot is $140\fdg3 \pm 0\fdg6$; the position angle of the quasar
from the hotspot location is a nearly identical 140\fdg8, indicating
that the radio lobe emanates from the quasar and is not a separate
source.  The southeast lobe is not reliably detected in the A-array
data, probably because it is too diffuse.

We observed J1016+5209 yet again with the VLA in 1999 November using
the B-configuration at a wavelength of 3.6~cm, this time obtaining
polarization information as well (Figure~4).  Because data were taken
at only one frequency, we are unable to correct for Faraday rotation;
however, since our measurements were at high frequency, this is
expected to be small since the angle of rotation $\theta \propto
\lambda^2$.  In fact, the orientation of the magnetic field lines is
as expected for a double-lobed FR-II source, parallel to the jet axis
until reaching the hotspots, where it becomes perpendicular to the
jet, indicating a shock-compressed field at the ends of the source.

Flux measurements at 8.46 GHz are: North hotspot/lobe = 18.6 mJy, Core
= 2.1 mJy ($\sim 4\%$ polarized with a position angle of $54\arcdeg
\pm 9\arcdeg$), South hotspot/lobe = 3.8 mJy.  The lobe fluxes are
lower limits as there will be some flux missing from the map on scales
$> 10\arcsec$.  The core is strongly polarized, 4\% at 8.4 GHz,
perhaps greater if there is any depolarization.  If the rotation
measure is high, which could be the case if J1016+5209 is embedded in
a thick shroud (see \S 3.1), depolarization of the core radio source could
be significant.  Comparing these flux density estimates to the FIRST
survey numbers, the spectral indices are -1.10, -0.63, and -1.31 for
the North lobe, Core, and South lobe, respectively.

\section {Analysis and Discussion}

The observed and derived properties of J1016+5209 are summarized in
Table~1; we adopt H$_{\circ}=50$~\kmsmpc\ and q$_{\circ}=0.5$.  The
radio properties are extreme in several respects: J1016+5209 has a
total radio luminosity of $10^{34.4}$ ergs cm$^{-1}$s$^{-1}$Hz$^{-1}$,
the highest of any known BAL quasar.
Using the definition of Stocke et al.\ (1992), we compute log(R$^*$),
the ratio of radio-to-optical brightness, as 3.4, using the global
radio spectral index of -1.1 and optical spectral index of -1.  This
value of log(R$^*$) is also an extreme for known BAL quasars, and is
at the high end of the distribution even for radio-selected non-BAL
quasars (White et al.\ 2000).  In fact, {\em even if we consider only
the core radio flux of J1016+5209, it is still radio-loud} with
log(R$^*$) = 2.0.

It is possible to estimate the angle between our line of sight and the
jet axis in J1016+5209 using the ``core-dominance'' measure defined by
Wills \& Brotherton (1995) as $\log(\mathcal{R}) = \log(L_R) +
0.4*M_{V} - 13.69$, where $L_R$ is the 5~GHz radio luminosity (of the
core alone, Table~1) in units of W~Hz$^{-1}$.  The $M_B = -26.2$
(Table~1) (we have assumed the $B$ and $O$ passbands to be
equivalent); we adopt $B-V = 1$ as a reasonable estimate of its color
based on the $O-E$ of 1.6.  This yields $\log(\mathcal{R}) \approx
1.3$, placing it at the extreme of orientations in the Wills \&
Brotherton sample where viewing angles are large, $>40\arcdeg$, but
not well constrained.  Even so, it is additional evidence that
J1016+5209 is viewed well away from the jet axis.

\subsection{The orientation model cannot explain BAL quasars}

A popular explanation for the presence of BALs in a minority ($\sim 10\%$)
of the quasar population is that quasars must be viewed at a preferred
orientation to exhibit BALs in their spectra, along a line of sight
roughly in the plane of the quasar accretion disk.  To test
this hypothesis, it is necessary to establish the orientation of BAL
quasars; this can be done through radio observations.  The FBQS is
finding a surprisingly large number of BAL quasars for a
radio-selected sample (Becker et al.\ 2000) with a frequency at least
as great as that of the optically selected Large Bright Quasar Survey
(Hewett, Foltz, \& Chafee 1995).  A large sample of radio-emitting BAL
quasars offers the chance to investigate BAL viewing orientations.

The present sample of $\sim 25$ FBQS BAL quasars consists
predominantly of unresolved compact radio sources, and, even the few
that are slightly resolved still show no outright structure (such as
lobes) on the angular scale of the FIRST survey (a few arcseconds).
This differs markedly from the radio morphologies of a non-BAL
subsample from the FBQS, matched in redshift and radio flux, where
30\% of the non-BAL quasars show extended radio structure (Becker et
al.\ 2000).  Even without direct evidence from the radio morphology,
some information on orientation can be obtained from radio spectra.
The radio spectral indices of the 29 BALs in the FBQS vary widely,
from -1.2 to +0.7, with 8 flatter than -0.5, and several more with
indices of -0.5 (Becker et al.\ 2000).  The scatter in spectral index
is consistent with the findings of Barvainis \& Lonsdale (1997) for
the radio spectral indices of a smaller sample of 15 BAL quasars.  If
the bulk of the radio emission is from a relativistic jet, the objects
with flat spectra are naturally interpreted as viewed close to the
radio jet axis, while the steep-spectrum objects are those seen at
larger viewing angles to the jet axis.  This analysis is at odds with
the model in which quasars must be viewed at a particular orientation
to see BALs (e.g., Weymann, et al.\ 1991; Hines \& Wills 1995;
Goodrich \& Miller 1995; Cohen et al.\ 1995).

Our results for J1016+5209 support the conclusion that a preferred
viewing angle is not necessary to produce a BAL quasar.  This object
has unambiguous properties of both a BAL and a radio-loud FR-II.  The
radio spectral indices of the core and lobes exhibit the expected
behavior, with the core flatter than the lobes, and the orientation is
clearly well away from the jet axis, quite the opposite from the
flat-spectrum BAL quasars.  We conclude that a preferred line of sight
is not necessary to observe BALs in quasars, and suggest that the
alternative view that the BAL phenomenon is a stage, probably early,
in the development of the quasar is more consistent with our data.
Becker et al.\ (2000) speculated that the compact radio size of the
BAL quasars implied a relatively young age for the BAL phenomenon: any
radio jet is still in the process of emerging from a thick cocoon of
material, and extended radio lobes on the 100 kpc scale have not had
time to develop, as has been suggested by Briggs, Turnshek, \& Wolfe
(1984), Voit et al.\ (1993), Egami et al.\ (1996), and others.  This
suggests a possible link to compact steep spectrum (CSS) or
gigahertz-peaked Spectrum (GPS) radio sources (e.g., O'Dea 1998; also
see below, \S 3.2).

Hamann et al.\ (1993) argue from detailed modeling of BAL spectra that
the covering factor $q$, the fraction of the sky covered by BAL
regions as seen from the central source, is $\sim 0.1$.  Coupled with
the statistical result that $\sim 10\%$ of all quasars have BALs,
their modeling result strongly implies that the conditions which give
rise to the BAL phenomenon are present in every quasar and we simply
do not see it 90\% of the time because our line of sight does not pass
through the BAL region.  One possible way to reconcile this with the
conclusion drawn above -- that BAL quasars are {\em not} seen at any
particular orientation -- is to relax the usual assumption that BAL
clouds are spatially concentrated near the plane of the obscuring
torus surrounding the central engine.  Arranging this could be
problematic, however, if the BAL clouds originate in the torus region
and are accelerated radially outward, which would naturally work to
confine them to the plane of the torus.

Perhaps a more likely explanation of the radio results for the FBQS
sample, which imply covering factors of approximately unity, is that
the critical assumption of the Hamann et al.\ models that photons are
conserved is not applicable.  Voit et al.\ (1993) argue that BAL
quasars which have very weak or absent \CIV\ emission cannot be
plausibly explained by small covering factors.  Rather, the \CIV\
photons are destroyed by repeated scatterings during their passage
through a spherical shell of gas and dust.  Such a shell does not even
need to be very optically thick to reduce the \CIV\ and other
resonance emission lines such as \MgII\ to a negligible amount, as
long as the dust and \CIV\ ions are co-spatial.  J1016+5209 certainly
has weak resonance emission lines and its continuum is quite red for a
quasar.  Adopting the ``starburst'' reddening law of Calzetti et al.\
(1994), we estimate that J1016+5209 has A$_{\rm V} \approx 0.75$, by
comparing the shape of the rest frame continuum between 1600\AA\ and
2200\AA\ with that of a composite quasar spectrum (Brotherton et al.\
2000) from the FBQS.  Figure~5 displays the observed and dereddened
spectrum of J1016+5209 and the quasar composite.  The derived A$_{\rm
V}$ implies an optical depth from dust at 1550\AA\ of $\sim 1.6$.  In
the simple scattering/absorption model that Voit et al.\ propose, this
is sufficient to destroy the large majority of resonance line photons
and provides a covering factor $\sim 1$.

The significant reddening from dust in J1016+5209 has further
implications.  First, high reddening is generally associated with BAL
quasars which show absorption from low ionization species such as
\MgII~2800, the ``LoBAL'' quasars (Sprayberry \& Foltz 1992).  Such
objects generally also have strong \FeII\ emission (Weymann et al.\
1991).  Inspection of Figures~1 and 5 reveals that the \MgII\ emission
is certainly much weaker in J1016+5209 than in the composite quasar,
perhaps because of weak BALs.  There is also a noticeable enhancement
of \FeII\ emission to the blue of \MgII, which may in fact be partly
responsible for filling in any possible \MgII\ BALs.  In the higher
resolution Keck spectrum of J1016+5209, there are weak absorption
lines which correspond to \AlIII\ $\lambda\lambda1854.7, 1862.8$.  The
\CIII\ emission feature in J1016+5209 is broader and not as peaked as
that of the composite quasar; this can be attributed to emission from
\FeIII\ $\lambda\lambda1895, 1926$.  We conclude that J1016+5209 has
some properties in common with LoBAL quasars.

All of these considerations again lead us to prefer a picture where
BAL quasars are emerging from a dusty cocoon of material, probably at
an early phase in their history.  The statistic that BAL quasars make
up 10\% of the quasar population suggests that this phase lasts about
10\% of the total quasar lifetime.  As LoBALs are generally more
highly reddened, they are an earlier period in the emergence of a
quasar in this model than HiBALs.

Correcting for the dust extinction makes J1016+5209 brighter at B by
1.1~magnitude, and so reduces its radio-loudness from log(R$^*$) = 3.4
to 3.0.  This still leaves it as the most radio-loud BAL known.

\subsection{J1016+5209 as a Transition or Hybrid Object}

J1016+5209 is the only FR-II quasar among the $\sim 50$ BAL quasars
which have been discovered in follow-up to the FIRST survey ($\sim 25$
from Becker et al.\ 2000, plus an additional $\sim 25$ in subsequent
follow-up, unpublished).  In the FBQS, $\sim 12\%$ of z$>0.5$ quasars
exhibit double-lobe morphology (Becker et al.\ 2000), and an
additional $\sim 10\%$ show at least some radio structure.  Why are
BAL quasars with large radio lobes so rare?  One possibility is that
J1016+5209, and its potential low-z counterpart PKS~1004+13, are
transition objects, on the way to becoming normal (non-BAL) FR-II
quasars, caught in a relatively brief period during which the two
phases co-exist.  Another possibility is that J1016+5209 is a hybrid
object, perhaps an old FR-II source which has recently been
rejuvenated as a CSS/BAL source in its core.

Even though the high resolution ESI spectrum shows that the BAL
features of J1016+5209 in general do not break up into myriad
cloudlets, the lowest velocity trough does exhibit more structure in
its depths than does the trough at -15000 \kms, and at one location,
v=-1400 \kms, a narrow inter-cloud continuum is nearly resolved
(Figure~3).  The absorption within 9000 \kms\ of the emission line
redshift is reminiscent of the more extreme examples of the class of
``associated absorber'' (AA) quasars (Foltz 1987).  Were it not for
its prominent BAL trough at -15000 \kms, J1016+5209 might fall more
naturally into the AA class, though it would be by far the most
extreme example.  A somewhat similar AA is PKS~1157+014 (Wright et
al.\ 1979), a z=1.9, radio-loud quasar.  It has two moderately broad
absorption troughs at -6500 and 0 \kms, not unlike the corresponding
but more extreme spectral regions of J1016+5209, even though BI=0 for
PKS~1157+014.  Whether such AA features, which are not broad enough to
gain distinction as true BALs in the quantitative BALnicity definition
of Weymann et al.\ (1991), are intrinsic to the quasar or generated in
an intervening object has been debated for some time (Morris et al.\
1986; Foltz et al.\ 1986).  Recently, Aldcroft, Bechtold, \& Foltz
(1998) presented evidence for variability of the higher velocity
outflow in PKS1157+014, perhaps resolving the argument in favor of the
intrinsic case, at least for this well-studied example.  This is
consistent with the growing evidence that many systems which are
currently thought to be intervening, especially in radio-loud quasars,
are really intrinsic (Richards et al.\ 1999).

It may be that J1016+5209 is in transition from a BAL to a more normal
radio-loud quasar and PKS1157+014 is representative of the next
evolutionary phase of a radio-loud object such as J1016+5209.  If the
highest velocity absorber in J1016+5209, already not as deep, were to
fade away first, leaving behind the lower velocity troughs, the result
would be similar to PKS1157+014.  Perhaps we have just happened to
catch J1016+5209 in a relatively rare, short-lived state in which it
exhibits BAL features while having already developed strong radio
emission.  This could occur in a brief period at the end of the
evolution of a BAL in which the radio emission finally manages to
erupt from confinement but the dense cocoon has not completely
dissipated.  During such dissipation, the BALs may eventually evolve
into distinct cloudlets as hinted at here, driven by the ensuing
outflows of ionized plasma accompanying the radio emission.  That the
central region of J1016+5209 is completely surrounded by turbulent
absorbing material is supported by the presence of absorption
occurring at velocities to the red of the rest frame \SiIV\ and \CIV\
by $\sim 1000$ \kms.

This picture is consistent with the radio core of J1016+5209 having an
unusually steep spectrum, $\alpha = -0.63$, and being unresolved at
the 0\farcs25 ($\sim 2$~kpc) scale.  These properties are reminiscent
of CSS or GPS sources: the leading interpretation of CSS and GPS
sources is that they are young radio objects, confined to a small
region by dense gas but which evolve with time into extended radio
sources with lobes as they escape confinement (O'Dea 1998), much like
the picture of BAL quasars emerging from cocoons (Voit et al.\ 1993).
This coincidence of attributes in J1016+5209 supports the notion that
BAL quasars are an early evolutionary phase in the life cycle of a
quasar.  The polarization of CSS sources is typically higher than that
of GPS sources, $\sim 5\%\ vs.\ 0.2\%$ at 6cm (O'Dea 1998), so the
$\sim 4\%$ polarization at 3.6cm for the core of J1016+5209 suggests
that it is a CSS object, but low frequency data for the core alone are
needed to confirm this.  If it is a CSS, then J1016+5209 may exhibit
the so-called ``alignment effect'' between its radio and optical
structure (McCarthy 1993 and references therein); any possible
connection with its BAL nature would be interesting in this context.
The misalignment of the optical polarization and large-scale radio jet
axes could be explained if on subarcsecond scales J1016-5209 has been
reborn with a different jet axis.

Steep-spectrum cores, however, are not uncommon in high redshift,
lobe-dominated quasars (Lonsdale, Barthel, \& Miley 1993).  An
alternative possibility is that J1016+5209 is a normal FR-II quasar in
a very low density environment which allows rapid expansion of the
radio lobes.  If the radio lobes were expanding unimpeded at
relativistic speeds, then the brighter, jet-side lobe should be
significantly farther from the core, whereas just the opposite is
seen.  The arm-length ratio for J1016+5209, however, is $Q = 0.64$, at
the extreme low end of the distribution found by Scheuer (1995) for a
sample of radio-luminous, double-lobed quasars.  The asymmetry of
J1016+5209 then is probably due to environmental rather than
relativistic effects, implying that the lobes are not expanding freely
and rapidly, and hence are not particularly young.  If the radio
source is expanding at speeds typical of FR-II radio sources, $\sim
0.1c$ (Arshakian \& Longair 2000), the large extent of the lobes of
J1016+5209, $\approx 350$ kpc, suggests a fairly advanced age (for a
radio source) of $\sim 10^7$ yr; PKS 1004+413 is also a large source,
$\approx 475$kpc in size, and so of comparable age.

If the core of J1016+5209 is a CSS or GPS object, the presence of
larger scale, presumably older, very bright radio lobes (Figures 3 and
4) at a large distance from the central engine supports the hypothesis
that quasars can be ``reborn'' and that perhaps both the BAL and
CSS/GPS properties can occur repeatedly in a given object, but always
early in any ``on'' cycle of AGN activity.  In support of the
rejuvenation picture, about 10\% of GPS/CSS sources have extended
emission (O'Dea 1998 and references therein), possibly from an earlier
period of activity, now dissipated.  It may be that J1016+5209 is in
an early phase of rejuvenation, having particularly compact inner
lobes which will grow, becoming a double-double radio source; a number
of such objects are known (Schoenmakers et al.\ 2000).  Higher
resolution mapping of the core of the J1016+5209 is needed to test its
GPS/CSS nature.

The interpretation of J1016+5209 as a rejuvenated quasar suggests that
it may not be correct to compute its radio-loudness using the entire
radio flux, at least not in the context of evaluating the
``BAL-related'' radio-loudness in its present incarnation.  With the
reddening correction and counting only the core flux, log(R$^*)
\approx 1.6$, still formally radio-loud, but not as exceptional as
$\gtrsim 3$ for the total radio flux.

\section {Summary}

The properties of the quasar FIRST J1016+5209 stand out in several
respects.  It exhibits bona fide BALs in its optical spectrum while
also having a classic FR-II radio-loud morphology.  J1016+5209 is the
most radio-loud and radio-luminous BAL quasar known.  The only other
object which may be of a similar nature is the less extreme, low
redshift quasar PKS~1004+13 (Wills et al.\ 1999).  
The presence of distinct bright radio lobes and its low ``core
dominance'' parameter implies that J1016+5209 is viewed well away from
the jet axis, at an angle of $\gtrsim 40^\circ$.  Based on the large
scatter in radio spectral indices, Becker et al.\ (2000) argue that
BAL quasars are not viewed at any particular orientation, contrary to
the popular orientation model.  The relatively steep spectrum ($\alpha
\approx -0.6$) and compact size ($< 0\arcsec3$ at 3.6cm) of the radio
core of J1016+5209 suggest that it is a CSS source, suggesting that
J1016+5209 is young.  This supports the alternate model of BAL quasars
in which they are an early phase in the evolution of quasars.  The
20\arcdeg--30\arcdeg\ misalignment of the optical and radio
polarization axes is further evidence that J1016+5209 does not easily
fit the orientation model for BAL quasars.

The large scale (350~kpc) FR-II radio lobes of J1016+5209 do not
easily fit the picture of it being young, so we postulate that it is a
rejuvenated quasar, possibly through a merger or interaction.  
If there is a newly created -- perhaps even episodic -- CSS source at the
core of J1016+5209, higher resolution imaging at various wavelengths
should reveal interesting connections among the various attributes
(BALs, CSS, radio-loudness, FR-II morphology) that have come together
in this one object.

\acknowledgments

We gratefully acknowledge D.~Stern and H.~Spinrad for obtaining the
low resolution Keck spectrum of FIRST J1016+5209 and Willem De Vries
for helpful comments.  Some of the data presented here were obtained
at the W.M.  Keck Observatory, which is operated as a scientific
partnership among the California Institute of Technology, the
University of California and the National Aeronautics and Space
Administration.  The Keck Observatory was made possible by the
generous financial support of the W.M. Keck Foundation.  The FIRST
Survey is supported by grants from the National Science Foundation
(grant AST-98-02791), NATO, the National Geographic Society, Sun
Microsystems, and Columbia University.  Part of the work reported here
was done at the Institute of Geophysics and Planetary Physics, under
the auspices of the U.S. Department of Energy by Lawrence Livermore
National Laboratory under contract No.~W-7405-Eng-48.

\pagebreak

\begin {deluxetable}{lr}
\tabletypesize{\small}
\tablewidth{4in}
\tablehead{\multicolumn{2}{c}{FIRST J101614.3+520916 Properties}\\
\colhead{} &
\colhead{}
}
\startdata
RA (J2000) &  $\mathrm{ 10^{h} 16^{m} 14\fs3 }$  \\
DEC (J2000) &  +52\arcdeg 09\arcmin 16\\
z   &         2.455 \\
$O (\approx B$)&       20.2 \\
$E (\approx R$)&       18.6 \\
S$_{\rm 20cm}$ core (mJy)&  6.5\tablenotemark{a} \\
S$_{\rm 20cm}$ NW lobe (mJy)  &  131.1\tablenotemark{a} \\
S$_{\rm 20cm}$ SE lobe (mJy) &   39.3\tablenotemark{a} \\
S$_{\rm 92cm}$ (mJy) &    850\tablenotemark{b} \\
S$_{\rm 6cm}$ (mJy) &    44\tablenotemark{c} \\
Lobe-lobe axis PA (\arcdeg)    &    146 \\
Lobe peak-to-peak distance (\arcsec) & 45\\
Total projected size (kpc) & 350 \\
M$_B$    &      -26.2 (-27.3)\\
log(L$_{5{\rm GHz}}$)~Total~(ergs s$^{-1}$cm$^{-2}$Hz$^{-1}$) & 34.3 \\
log(L$_{5{\rm GHz}}$)~Core~(ergs s$^{-1}$cm$^{-2}$Hz$^{-1}$) & 32.9 \\
log(R$^*$) (total) &     3.4 (3.0) \\
log(R$^*$) (core only) &     2.0 (1.6) \\
\enddata
\tablenotetext{a} {FIRST survey (Becker et al.\ 1995)}
\tablenotetext{b} {WENSS (Rengelink et al.\ 1997)}
\tablenotetext{c} {Greenbank (Gregory et al.\ 1996)}
\tablecomments{Derived quantities assume H$_{\circ}=50$~\kmsmpc\ and q$_{\circ}=0.5$.
Quantities in parentheses corrected for intrinsic
extinction of A$_{\rm B}=1.1$}
\end {deluxetable}

\pagebreak

\begin{figure}
\plotone{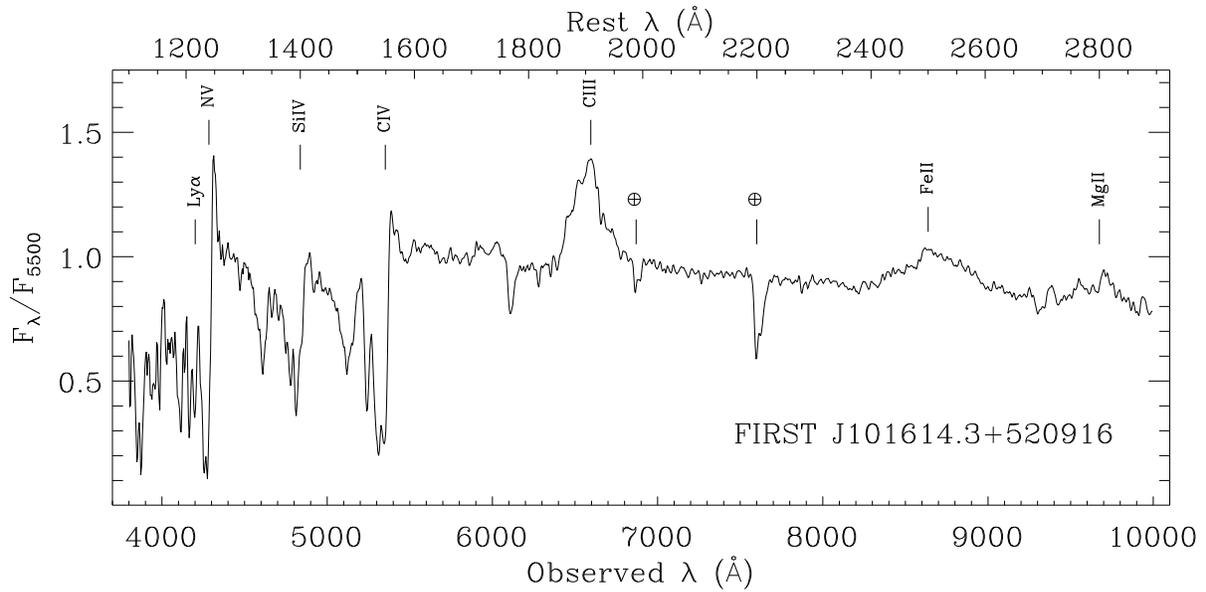}
\caption{Low resolution spectrum of J101614.3+520916 obtained at Keck
Observatory using LRIS.  Prominent features are labeled.  There is a
relatively strong intervening \MgII\ 2800 absorption system 
at an observed wavelength of 6110\AA\ (z = 1.182).}
\end{figure}

\begin{figure}
\epsscale{0.7}
\plotone{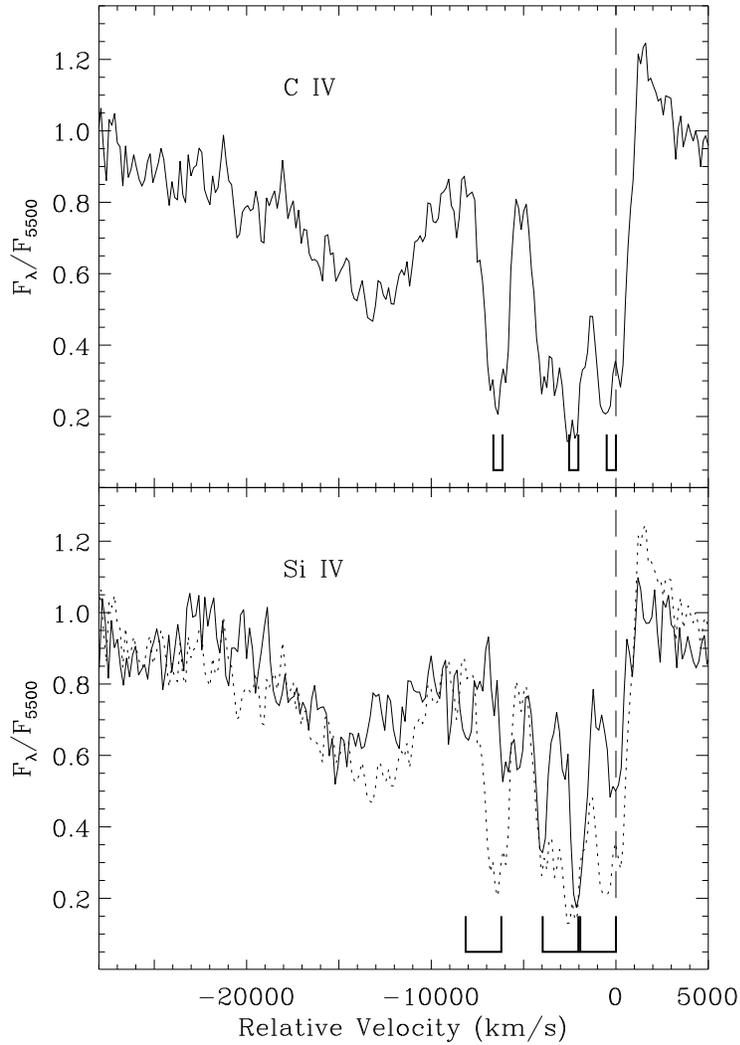}
\caption{Close-up of the \CIV\ and \SiIV\ regions of the spectrum of FIRST
J101614.3+520916, shown in velocity space appropriate for the rest
frame of the red side
of each doublet.  The \CIV\ region is repeated
in the the lower panel as a dotted line for comparison.}
\end{figure}

\begin{figure}
\epsscale{1.}
\plotone{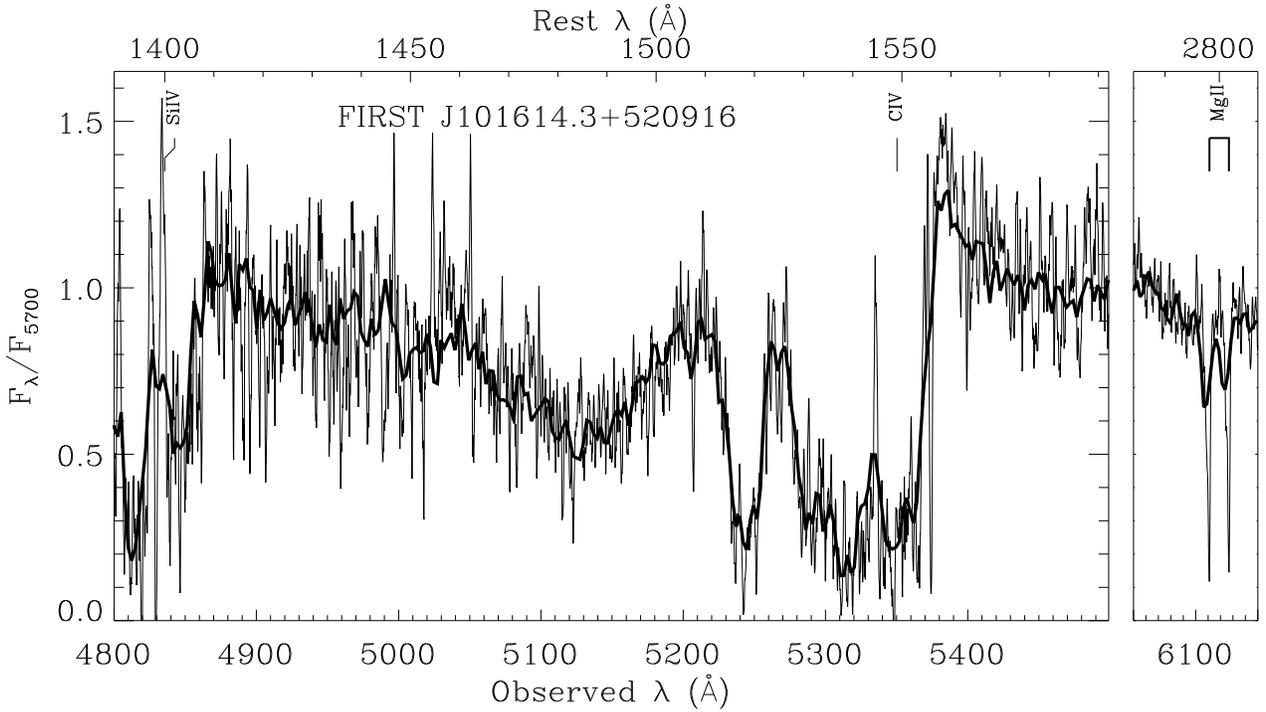}
\caption{Keck 10m ESI spectrum of the \CIV\ BAL region (thin line) in
J101614.3+520916; the resolution is 1.5\AA.  Overplotted is the 9\AA\
resolution LRIS spectrum (thick line) from Figure~2.  The gain in
resolution with ESI compared to LRIS is demonstrated by the small panel
on the right showing the intervening \MgII\ absorber in each.  The
\CIV\ absorption troughs do not break up into numerous individual
unresolved cloudlets, confirming the BAL nature of J1016+5209.}
\end{figure}

\begin{figure}
\epsscale{1.}
\plotone{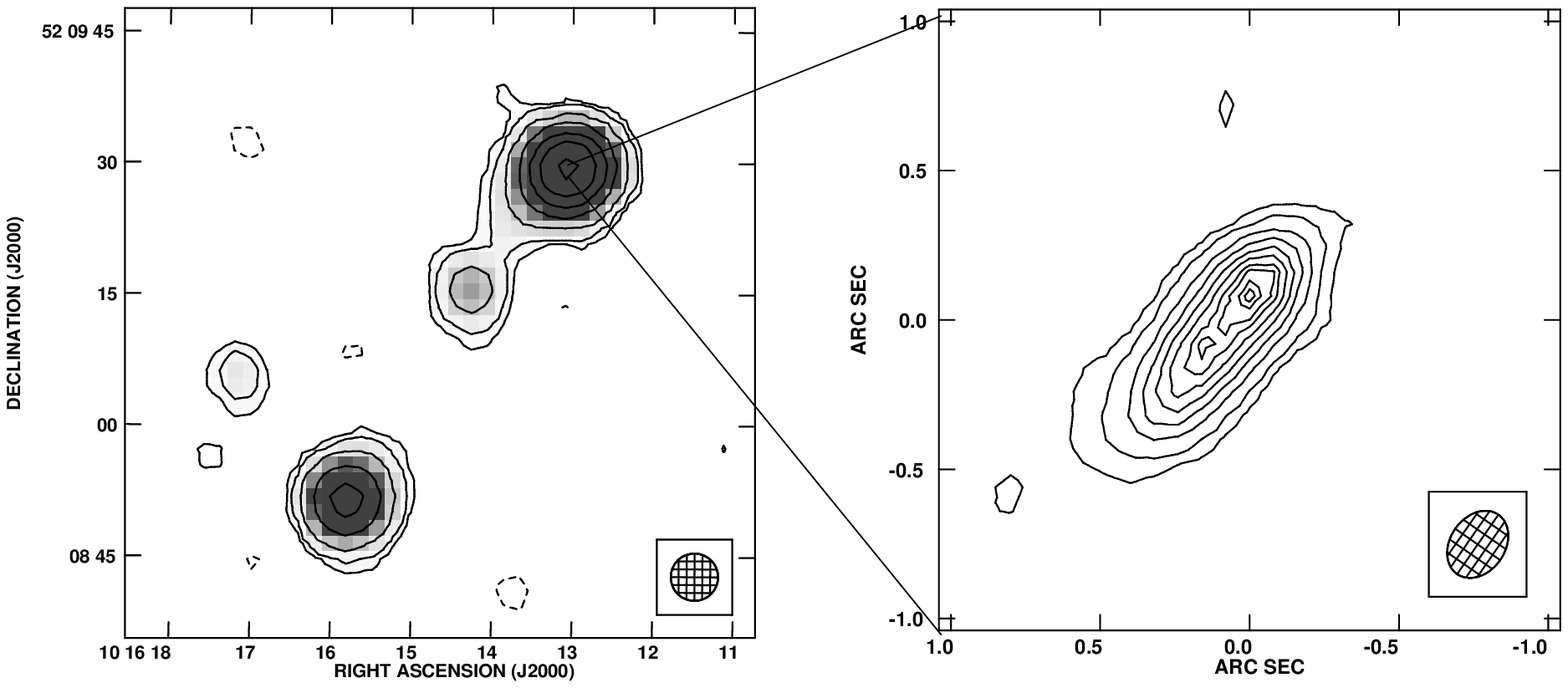}
\caption{Left: FIRST survey 20~cm image of J101614.3+520916; contour
levels are -0.5, 0.5, 1.0, 5, 10.0, 20.0, 50.0 and 100.0 mJy.\newline
Right: A-array 3.6~cm image of the northwest lobe showing linear
structure pointing directly back to the radio core, evidence that the
two are physically associated and not a chance arrangement of radio
sources.  Contour levels are 0.24, 0.5, 1.0, 1.5, 2.0, 2.5, 3.0, 3.35,
3.75, and 4.0 mJy; the map has an RMS of 0.078~mJy.}
\end{figure}

\begin{figure}
\epsscale{0.7}
\plotone{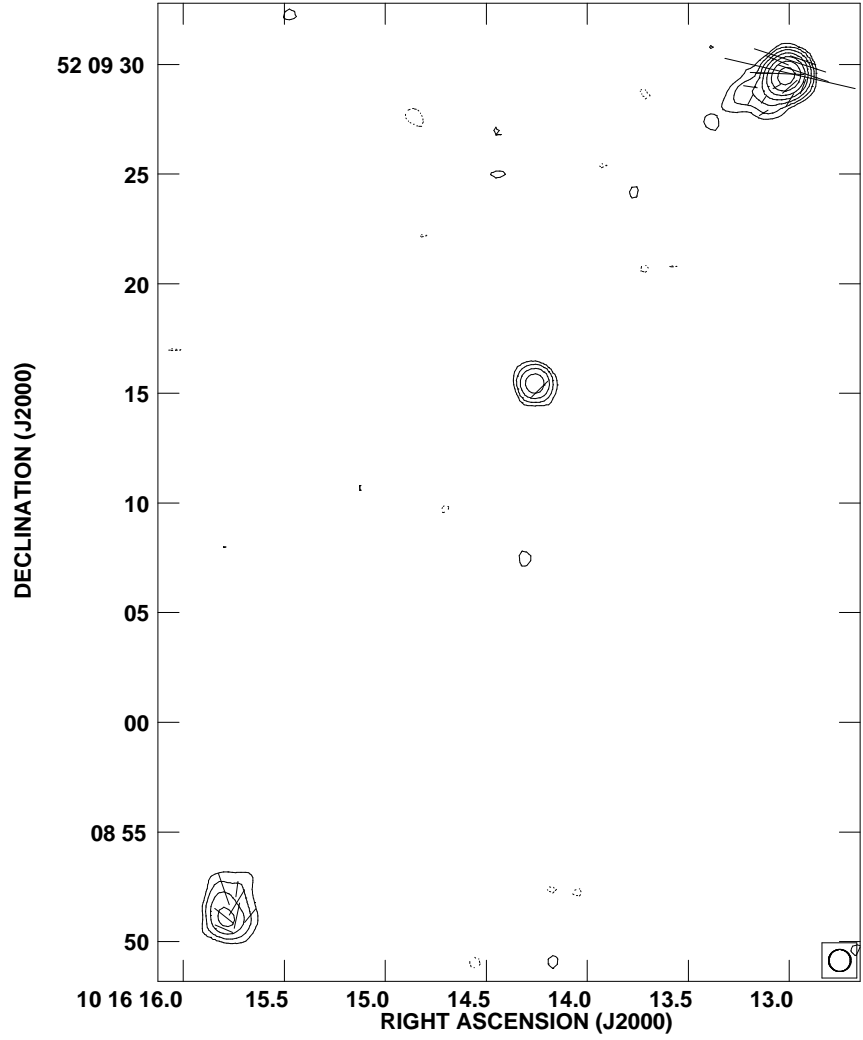}
\caption{VLA B-array 3.6cm map of J101614.3+520916; contours are
logarithmic, spaced by factors of two from $\pm 0.15$ mJy.  The
polarization vectors have been rotated by 90\arcdeg\ to indicate the
orientation of the magnetic field and are scaled so that a polarized
flux of 0.1 mJy corresponds to a vector one arcsecond long.  No
corrections for Faraday rotation or depolarization have been made.
Flux measurements at 8.46 GHz are: North hotspot/lobe = 18.6 mJy, Core
= 2.1 mJy, South hotspot/lobe = 3.8 mJy.  }
\end{figure}

\begin{figure}
\epsscale{1.}
\plotone{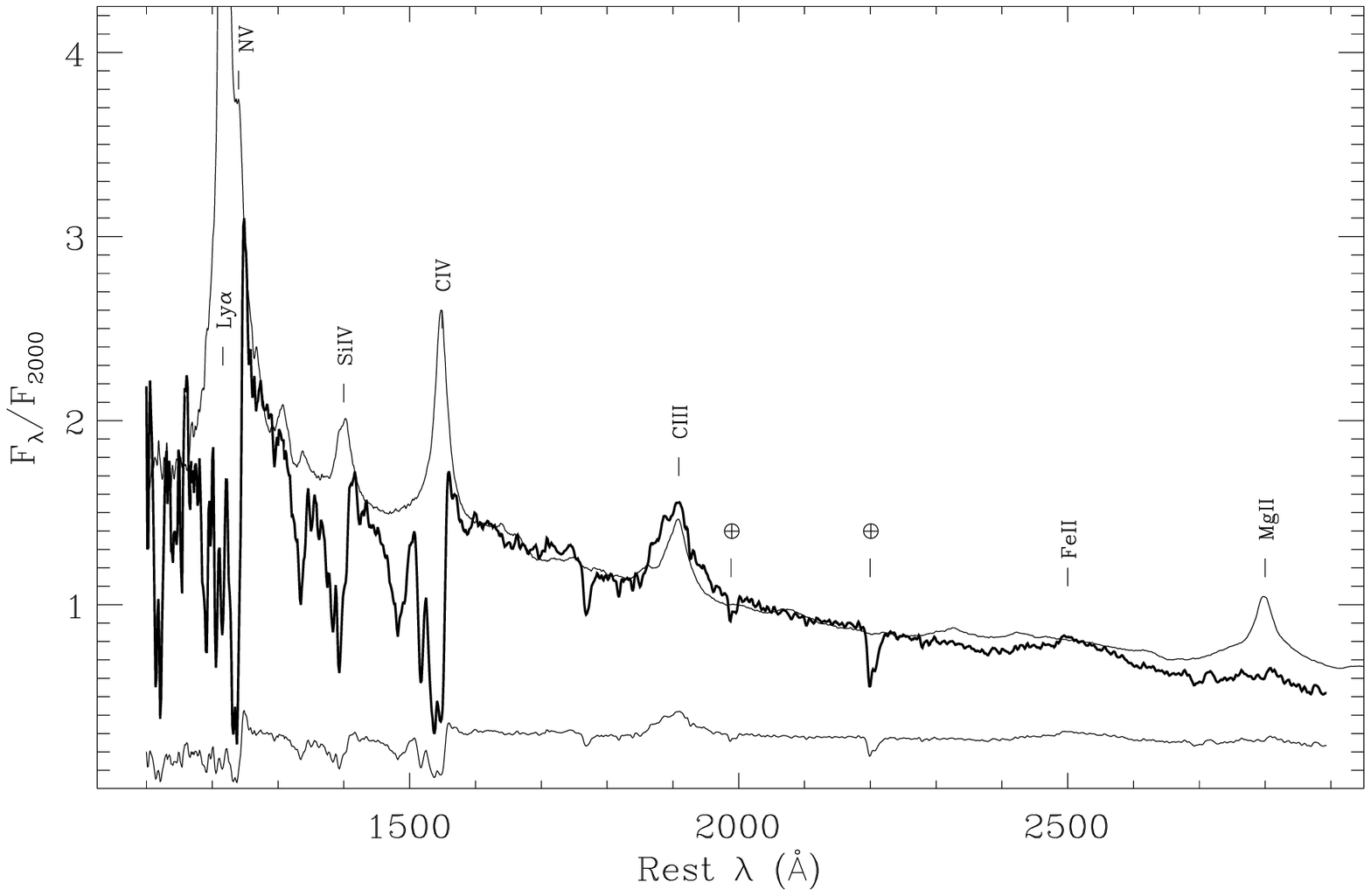}
\caption{Dereddening the Keck LRIS spectrum of FIRST J1016+5209 (lower
thin line) using A$_{\rm V} = 0.75$ and the starburst reddening law of
Calzetti et al.\ (1994) yields a good match (thick line) with the
composite quasar spectrum (Brotherton et al.\ 2000) from FBQS (upper
thin line).  Prominent features are labeled; J1016+5209 has
comparatively weak \MgII~2800, suggesting possible absorption by BALs,
in turn possibly masked by \FeII\ emission which is common in LoBALs.
The \CIII~1909 emission profile also appears to be distorted relative
to the composite; this is likely from
Fe~{\sc III}, Si~{\sc III}, and Al~{\sc III} emission.
}
\end{figure}


\begin{references}

\reference{} Arshakian, T. G. \& Longair, M. S. 2000, MNRAS, in press

\reference{} Barvainis, R., \& Lonsdale, C. 1997, AJ, 113, 144

\reference{} Becker, R. H., Gregg, M. D., Hook, I. M., McMahon, R. G.,
White, R. L.,
\& Helfand, D. J.  1997, ApJL, 479, L93

\reference{} Becker, R. H., White, R. L., \& Edwards, A. L. 1991, ApJS, 75, 1

\reference{} Becker, R. H., White, R. L., Gregg, M. D., Brotherton, M.S.,
Laurent-Muehleisen, S. A., Arav, N. 2000, ApJ, in press

\reference{first} Becker, R. H., White, R. L., \& Helfand, D. J. 1995, ApJ,
450, 559

\reference{} Begelman, M. C., de Kool, M., \& Sikora, M. 1991, 382, 416

\reference{} Boroson, T. A., Persson, S. E. \& Oke, J. B. 1985, 293, 120

\reference{} Briggs, F. H., Turnshek, D. A., \& Wolfe, A. M. 1984, ApJ,
287, 549

\reference{} Brotherton, M. S., van Breugel, W., Smith, R. J., Boyle,
B. J., Shanks, T., Croom, S. M., Miller, L. \& Becker, R. H.  1998, ApJL,
505, L7

\reference{} Brotherton, M. S., Tran, H. D., Becker, Gregg, M. D., 
R. H., Laurent-Muehleisen, S. A., \& White, R. L. 2000, in preparation

\reference{} Calzetti, D., Kinney, A. L., Storchi-Bergmann, T. 1994,
ApJ, 429, 582

\reference{} Cohen, M. H., Ogle, P. M., Tran, H. D., Vermeulen, R. C.,
Miller, J. S., Goodrich, R. W., \& Martel, A. R. 1995, ApJL, 448, L77

\reference{} Condon, J. J., Cotton, W. D., Greisen, E. W., Yin, Q. F.,
Perley, R. A., Taylor, G. B. \& Broderick, J. J. 1998, AJ, 115, 1693

\reference{} Epps, H. W. \& Miller, J. S. 1998, Proc.\ SPIE
3355, Optical Astronomical Instrumentation, edited by Sandro
D'Odorico, p.~48

\reference{} Egami, E., Iwamuro, F., Maihara, T., Oya, S., \& Cowie,
L. L. 1996, AJ, 112, 73

\reference{} Fanaroff, B. and Riley, J. M., 1974, MNRAS, 167, 31P

\reference{} Foltz, C. B., Weymann, R. J., Peterson, B. M., Sun, L.,
Malkan, M. A., \& Chaffee, F. H., Jr.  1986, ApJ, 307, 504

\reference{} Foltz, C. B. 1988, in, QSO Absorption Lines: Probing the
Universe (Space Telescope Science Institute Symposium Series 2),
edited by J. C. Blades, D. A. Turnshek, and C. A. Norman (Cambridge
University Press, New York), p. xxx

\reference{} Goodrich, R. W. \& Miller, J. S. 1995, ApJL, 448, L73

\reference{} Gregg, M. D., Becker, R. H., White, R. L., Helfand, D. J., 
McMahon, R. G., \& Hook, I. M. 1996, (FBQS I), AJ, 112, 407

\reference{} Hamann, F., Korista, K. T., \& Morris, S. L. 1993, ApJ,
415, 541

\reference{} Hewett, P. C., Foltz, C. B., \& Chaffee, F. H. 1995,
AJ, 109, 1498

\reference{} Hines, D. C. \& Wills, B. J. 1995, ApJL, 448, L69

\reference{} Lonsdale, C. J., Barthel, P. D., \& Miley, G. K. 1993, ApJS,
87, 63

\reference{} McCarthy, P. J. 1993, ARAA, 31, 639

\reference{} Morris, S. L., Weymann, R. J., Foltz, C. B., Turnshek,
D. A., Shectman, S., Price, C., \& Boroson, T. A. 1986, ApJ, 310, 40

\reference{} O'Dea, C. P. 1998, PASP, 110, 493

\reference{} Oke, J. B., Cohen, J. G., Carr, M., Cromer, J., Dingizan,
A., Harris, F. H., Labrecque S., Lucinio, R., Schaal, W., Epps, H., \&
Miller, J. 1995, PASP, 107, 375

\reference{} Rengelink, R. B., Tang, Y., de Bruyn, A. G., Miley, G. K.,
Bremer, M. N., R\"ottgering, H. J. A., \& Bremer, M. A. R.
1997, A\&AS, 124, 259

\reference{} Richards, G. T., York, D. G., Yanny, B., Kollgaard, R. I.,
Laurent-Muehleisen, S. A., \& Vanden Berk, D. E., 1999, ApJ, 513, 576

\reference{} Scheuer, P. A. G. 1995, MNRAS, 277, 331

\reference{} Schoenmakers, A. P., de Bruyn, A. G., R\"{o}ttgering,
H. J. A., van der Laan, H., \& Kaiser, C. R. 2000, MNRAS, in press

\reference{} Sprayberry, D., \& Foltz, C. B. 1992, ApJ, 390, 39

\reference{} Stocke, J. T., Morris, S. L., Weymann, J. T., \& Foltz,
C. B. 1992, ApJ, 396, 487

\reference{} Voit, G. M., Weymann, R. J., \& Korista, K. T. 1993, ApJ,
95, 109

\reference{} Weymann, R. J., Morris, S. L., Foltz, C. B., \& Hewett, P. C.
1991, ApJ, 373, 23

\reference{} White, R. L., et al.\ (FBQS II) 2000, ApJS, in press

\reference{} Wills, B. J., Brandt, W. N., \& Laor, A. 1999, ApJL, 520, L91

\reference{} Wills, B. J. \& Brotherton, M. S. 1995, ApJL, 448, L81

\reference{} Wright, A. E., Morton, D. C., Peterson, B. A., \&
Jauncey, D. L. 1979, MNRAS, 189, 611

\end{references}
\end{document}